\documentstyle[sprocl,epsfig]{article}

\def\be{\begin{equation}}
\def\ee{\end{equation}}
\def\bea{\begin{eqnarray}}
\def\eea{\end{eqnarray}}

\newcommand{\nc}{\newcommand}
\nc{\del}{\partial}
\nc{\slsh}[1]{\!\!\not\!#1}
\nc{\slsho}[1]{\!\not\!\!#1}
\nc{\half}{\frac{1}{2}}
\nc{\non}{\nonumber}
\nc{\bm}[1]{\mbox{\boldmath $#1$}}
\nc{\bd}{\begin{displaymath}}
\nc{\ed}{\end{displaymath}}
\nc{\beas}{\begin{eqnarray*}}
\nc{\eeas}{\end{eqnarray*}}
\nc{\up}{\uparrow}
\nc{\dn}{\downarrow}
\nc{\ket}[1]{\left|#1\right>}
\nc{\bra}[1]{\left<#1\right|}
\newcounter{exercise}
\addtocounter{exercise}{1}
\nc{\ex}{\hspace*{10mm}\bm{{\rm Ex \,\char 35}\arabic{exercise}}\addtocounter{exercise}{1}}
\nc{\exv}{\arabic{exercise}}
\newcounter{example}
\addtocounter{example}{1}
\nc{\eg}[1]{\subsubsection{Example \,\char 35 \arabic{example} #1}\addtocounter{example}{1}}
\nc{\egv}{\arabic{example}}

\nc{\ru}{\rule{5cm}{0.2mm}\hfill\rule{5cm}{0.2mm}\vskip 2.5cm\rule{5cm}{0.2mm}\hfill\rule{5cm}{0.2mm}}

\begin{document}

\title{Classical Quark Models: \\
An Introduction}

\author{A. W. Thomas and S. V. Wright}

\address{Department of Physics and Mathematical Physics\\
and Special Research Center for the Subatomic Structure of Matter,\\
University of Adelaide, SA 5005, Australia\\
E-mail: athomas,swright@physics.adelaide.edu.au}


\maketitle\abstracts{
We present an elementary introduction to some of the quark models used
to understand the properties of light mesons and baryons. These lectures
are intended for both theoretical and experimental 
graduate students beginning their study of the strong interaction.
}

\vspace{-9cm}
\begin{flushright}
{\footnotesize Lectures at the XI Physics Summer School} \\
{\footnotesize Frontiers in Nuclear Physics \hspace{3cm}} \\
{\footnotesize ANU (Canberra), January 12-23, 1998} \\
{\footnotesize ADP-98-46/T316 \hspace{3cm}}
\end{flushright}
\vspace{7.8cm}

\section{Introduction}

Modern nuclear and particle physics has a secure theoretical foundation
based on the idea of local gauge invariance, known as the standard
model. It has satisfied every experimental or theoretical challenge 
directed at it so far. In the electroweak sector it was spectacularly
confirmed by the discovery of the heavy vector bosons (the $W^\pm$ and
$Z^0$) in the early 80's. On the other hand, in the strong interaction
sector, quantum chromodynamics (QCD) has presented more problems. While
it has passed every test in the regions where it can be solved (notably
the QCD evolution of parton distributions in deep inelastic scattering),
it has not been possible to solve it satisfactorily in the
non-perturbative sector, especially for the structure and properties of
the light hadrons, such as the nucleon.

The fundamental particles upon which QCD is built are the quarks. 
In the early 1960's there was a rapid proliferation in the number of
``elementary'' strongly interacting particles (hadrons) and it was 
suggested that these were more likely composite particles.  The more
{}fundamental particles from which they were built were initially 
called {\em Aces} or {\em Quarks}. By choosing them as a fundamental
representation of the unitary symmetry group SU(3), one could
immediately bring order to the chaos that had previously reigned in
hadron spectroscopy. One unsatisfactory feature of the quark hypothesis
was that no-one had observed a fractionally charged particle. 
They were sought in every conceivable place, from deep ocean sediments
to rocks brought back from the moon.  It
well known that none were ever detected -- at least as free particles.

On the other hand, there has been tremendous progress on the strong
interactions, both theoretically and experimentally. There are firm
indications that non-perturbative QCD will never allow us to see free
quarks, that is, that they are forever confined to the interior of
hadronic systems in a colour singlet state. Nevertheless, there is an
overwhelming amount of evidence to support the idea that quarks, with
precisely the expected electroweak properties, do exist in the interior
of the observed hadrons. Indeed, one of the most exciting challenges
facing nuclear physics is to investigate the role played by quarks in
explaining the properties of nuclear matter, from the normal densities of
observed nuclei through to the much higher densities at the centre of
neutron stars and the possible deconfinement transition in relativistic
heavy ion collisions.

Initially, only three types of quark were required to understand the
known hadrons, the {\em up}, {\em down} and {\em strange} quarks
($u,d,s$). However, the discovery of the $J/\psi$ particle in 1975 
increased this list to include the {\em charm} quark, $c$.  Later the 
discoveries of the $\Upsilon$ led to the inclusion of the {\em bottom/beauty}
quark, $b$, and finally the {\em top/truth} quark, $t$.

We shall be concerned with just the three light quarks which are of most
interest in nuclear physics. Moreover, because of the space limitations,
it is not possible to discuss all of the models which have been invented
to represent the non-perturbative regime of QCD. Over time these models
have tended to become more and more complicated and, in some cases very
abstract, so that the beauty and simplicity which led to the quark model
in the first place can often be lost. In these lectures we shall
concentrate on just a few, relatively simple models:
\begin{enumerate}
\item Non-relativistic Quark Models
	\begin{itemize} 
	\item centre of mass can be done exactly,
	\item chiral symmetry -- problems,
	\item relativity -- problems.
	\end{itemize}
\item Nature of confinement
	\begin{itemize} 
	\item vacuum structure,
	\item soliton.
	\end{itemize}
\item Relativistic, confining, potential model.
\item MIT Bag 
	\begin{itemize} 
	\item symmetries,
	\item conserved currents.
	\end{itemize} 
\item Chiral Bags.
\end{enumerate}
Furthermore, our presentation will be at a very elementary level. It is
intended primarily for beginning graduate students interested in the
strong interaction -- from both the experimental and theoretical
perspectives. Although set at a relatively elementary level, we hope that
the beginning student will find enough background, not often presented
now, to be able to better appreciate the current literature together
with a few insights that may help to make the subject come alive.

\section{Harmonic Oscillator Shell Model}
\label{sec:h_osc}

\begin{figure}[hbt]
\centering{\
	\epsfig{angle=0,figure=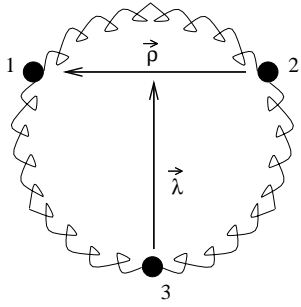, height=4cm} }
\parbox{100mm}{\caption{Picture of Harmonic Oscillator Shell Model.}
\label{fig:H.Osc}}
\end{figure}

The first model that we shall look at is the harmonic oscillator shell
model.  
This is quite a simple model, but it is nevertheless helpful in
understanding many features of the hadron spectrum. the model is built
on the valence quark picture in which a hadron consists
of just
three, confined, quarks.  The concept of confinement is represented by
linking the quarks together by virtual springs, so that a baryon, for
example, is described
mathematically by the following Hamiltonian
\be
H = \sum^{3}_{i=1}\frac{\vec{p}_{i}{}^{2}}{2m} + 
\half \kappa\sum_{i<j}\left|\vec{r}_{i} - \vec{r}_{j}\right|^{2}.
\label{eqn:h_osc_H}
\ee
The harmonic oscillator has the convenient property that the centre of
mass motion can be exactly separated from the internal dynamics. In
addition, the internal structure can be written as an effective two-body
problem,
with two quarks combined as a single subsystem. 
This involves the following definitions
\bea
\vec{\rho} = \frac{1}{\sqrt{2}}\left(\vec{r}_{1} - \vec{r}_{2}\right), \\
\vec{\lambda} = \frac{1}{\sqrt{6}}\left(\vec{r}_{1} + 
\vec{r}_{2} - 2\vec{r}_{3}\right),
\eea
and finally
\be
\vec{R} = \frac{\left(\vec{r}_{1}+\vec{r}_{2}+\vec{r}_{3}\right)}{3},
\ee
where the coordinates can be visualised as in Fig.\ \ref{fig:H.Osc}.

These definitions can be used to simplify the form of the Hamiltonian, 
Eq. (\ref{eqn:h_osc_H}), so it is easier to visualise the two body nature.
\be
H = H_{\rm CM} + \frac{\vec{p}_{\rho}{}^{2}}{2m} + 
\frac{\vec{p}_{\lambda}{}^{2}}{2m} + \frac{3\kappa}{2}\left(\rho^{2} + 
\lambda^{2}\right),
\ee
where $\vec{p}_{\rho} (\vec{p}_{\lambda})$ is the momentum conjugate 
to $\vec{\rho}$ ($\vec{\lambda}$) and $H_{\rm CM}$ is the kinetic energy of the
centre of mass motion of the three-quark system as a whole.
Clearly the three-body Hamiltonian has separated into two distinct
oscillators describing the internal structure of the hadron. In fact, in
this equal mass case, 
the oscillator frequencies of both two-body systems are the same
\be
\omega_{\rho} = \omega_{\lambda} = \sqrt{\frac{3\kappa}{m}} = \omega,
\ee
and defining a constant $\alpha$, as
\be
\alpha = \sqrt{m\omega},
\ee
we find that we can solve for the following wavefunctions.
\vspace{5mm}
\begin{center}
\begin{tabular}{|c|l|l|}
\hline
{\bf Quantum No.'s} & {\bf Wavefunction} & {\bf Energy} \\
\hline
$N = 0$ & $\psi_{00} = \frac{\alpha^{3}}{\pi^{3/2}}\exp^{-\alpha^{2}\left(\lambda^{2} + \rho^{2}\right)}$ & $E_{0} = \frac{3}{2}\hbar\omega + \frac{3}{2}\hbar\omega$ \\
\hline
$N = 0$ & $\psi_{11}^{\lambda} = \frac{\alpha^{4}}{\pi^{3/2}}\lambda{\cal Y}_{lm}(\hat{\lambda})\exp^{-\alpha^{2}\left(\lambda^{2} + \rho^{2}\right)}$ & $E_{1} = \frac{5}{2}\hbar\omega + \frac{3}{2}\hbar\omega$ \\
$(L = 1)$ & $\psi_{11}^{\rho} = \frac{\alpha^{4}}{\pi^{3/2}}\rho{\cal Y}_{lm}(\hat{\rho})\exp^{-\alpha^{2}\left(\lambda^{2} + \rho^{2}\right)}$ & $E_{1} = \frac{5}{2}\hbar\omega + \frac{3}{2}\hbar\omega$ \\
\hline
& $\;\;\;\;\;$etc... & \\
\hline
\end{tabular}
\end{center}
\vspace{5mm}

A physical model is of limited use if it cannot be compared with real
data. Within the standard model, the quark masses are currently free
parameters. In the oscillator model these are often set to the values
(constituent quark masses):
\be
\begin{array}{l}
m_{u} \simeq m_{d} \simeq 340 {\rm MeV}, \\
m_{s} \simeq 500 {\rm MeV}.
\end{array}
\ee
Having quarks of unequal mass slightly complicates the kinematics.
Suppose we have two quarks of equal mass and one different, then we
can set:
\be
m_{1} = m_{2} = m \; ; \; m_{3} = m^{\prime},
\ee
so that, through the definitions
\be
m_{\rho} = m \; ; \; m_{\lambda} = \frac{3mm^{\prime}}{(2m + m^{\prime})},
\ee
we can assign masses to the two body subsystems. Using these definitions
in the appropriate Hamiltonian we find that in the centre of mass system
($H_{\rm cm} = 0$):
\be
H = \frac{\vec{p}_{\rho}{}^{2}}{2m_{\rho}} + 
\frac{\vec{p}_{\lambda}{}^{2}}{2m_{\lambda}} + 
\frac{3\kappa}{2}\left(\rho^{2} + \lambda^{2}\right)
\ee
and we notice that $\omega_{\rho}\neq\omega_{\lambda}$.

As a simple consequence of the fact that
$\omega_{\rho}\neq\omega_{\lambda}$, as noted in the previous paragraph,
the excitation energies of the $\rho$ and $\lambda$ degrees of
freedom are different. Even this simple observation has profound
phenomenological consequences. One of the mysteries of the baryon
spectrum is that there are far fewer states observed than predicted by
the quark model. A possible explanation of this, advanced by Isgur and
Karl can be illustrated by the usual experimental method for exciting
baryon resonances. 
For example, in the $\bar{K}+N \rightarrow H \rightarrow \bar{K}+N$ reaction 
we find that the $\omega_{\lambda}$ degree of freedom 
tends to be excited rather than 
$\omega_{\rho}$. (This is a direct consequence of the different energies 
required to excite the strange and non-strange degrees of freedom.)  
As a result one will tend to be experimentally blind to those strange baryon
resonances where the excitation energy has gone into the $\omega_{\rho}$
coordinate. In order to test this idea one needs new ways of exciting
baryon resonances and this is precisely what is planned at new,
high-duty factor machines like CEBAF.
\begin{figure}[hbt]
\centering{\
	\epsfig{angle=0,figure=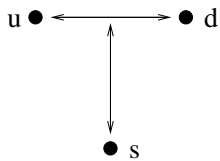, height=2cm} }
\parbox{100mm}{\caption{Idealised picture of, for instance, a $\Lambda$.}
\label{fig:b}}
\end{figure}

\section{Spectroscopy}

The hadron wavefunction is dependent upon the intrinsic properties of the 
quarks:
\bea
\Psi \equiv \Psi({\rm space}, {\rm spin}, {\rm flavour}, {\rm colour}). \non
\eea
{}For baryons it must be totally symmetric in the
first three of these, {\em space}, {\em spin}, and {\em flavour}, 
because the colour wave function must be 
totally anti-symmetric in order to produce a colour singlet object.
Taking the $\Delta$ as an example, we 
know that the {\em space}, {\em spin}, and {\em flavour} components of
the wave function are each symmetric ($S$),
\bea
\Delta: S, S, S \non
\eea
However for the $N$ each of the {\em spin} and {\em flavour} components
is of mixed symmetry ($MS$), with a symmetric {\em spatial} wave
function:   
\bea
N: S, MS, MS. \non
\eea
To first order the strong interaction we expect SU(6) spin-flavour symmetry
to be respected for the light quarks, so given one hadronic state we can
generate many others with raising and lowering operators.
Below we present an example (and some exercises for the reader) of how
operating on the wavefunction of one particle predicts the existence of 
another.  (The solutions to some of the 
exercises are included in the appendix.)

\eg{ -- Spin-Isospin}
Take for example the spin-isospin (SI) wavefunction of the $\Delta^{++}$
\be
\left|\Delta^{++},J_{z}=+\frac{3}{2}\right>_{\rm SI} = 
\left|u^{\uparrow}_{1},\:u^{\uparrow}_{2},\:u^{\uparrow}_{3}\right>.
\ee
We can calculate the wavefunction of the $\left|\Delta^{+},J_{z}
=+\frac{3}{2}\right>$ by operating on it with the lowering 
operator $I_{-} (\propto \tau_{1-} + \tau_{2-} + \tau_{3-})$ with
\be
\tau_{1-}\left|u_{1}\right> = \left|d_{1}\right>,
\ee
and
\be
\left[I_{-},I^{2}\right] = 0.
\ee
Therefore, after this operation we have
\be
\left|\Delta^{+},J_{z}=+\frac{3}{2}\right> 
= \frac{1}{\sqrt{3}}\left|\widetilde{u\uparrow,\:u\uparrow,\:d\uparrow}\right>,
\ee
where $\left|\widetilde{\;\;\;}\right> \equiv$ 
sum over all {\bf different} permutations.

Similarly, and this is left as an excise for the reader, we can see the 
change in the spin of the particle by operating with the spin 
lowering operator $S_{-}$,
\bea
\left|\Delta^{+},J_{z}=+\frac{1}{2}\right> & \propto & 
S_{-}\left|\Delta^{+},J_{z}=+\frac{3}{2}\right> \non \\
& = & \frac{1}{3}\left(\left|\widetilde{u\uparrow,\:u\uparrow,\:d\downarrow}
\right> + \left|\widetilde{u\uparrow,\:u\downarrow,\:d\uparrow}
\right>\right). \\
& & \ex \non
\label{eqn:ex_Delta+1/2}
\eea
As a hint for this exercise, the first wavefunction on the right 
hand side of the equality in Eq. (\ref{eqn:ex_Delta+1/2}) has 3 terms, and 
the second has 6 terms.  The spin-up proton is the orthogonal 
state with the same values of $I_3$ and $S_z$, and has wavefunction
\be
\left|p\uparrow\right>_{\rm SI} = 
\frac{1}{18}\left(2\left|\widetilde{u\uparrow,\:u\uparrow,\:d\downarrow}\right> 
- \left|\widetilde{u\uparrow,\:u\downarrow,\:d\uparrow}\right>\right). \ex
\ee
{}For {\bf Exercise \exv} find the SI wavefunction of 
$\left|\Sigma^{0}\uparrow\right>$ and $\left|\Lambda\uparrow\right>$. 
(Hint:  Start from $\left|\Sigma^{*+},J_{z}=+\frac{3}{2}\right> = 
\left|\widetilde{u\uparrow,\:u\uparrow,\:s\uparrow}\right>$).

\addtocounter{exercise}{1}
{\bf Exercise \exv:}  Non-relativistically the magnetic moment of a 
particle of charge $q$, and mass $m$ is
\addtocounter{exercise}{1}
\be
\vec{\mu} = \frac{q}{2m}\vec{\sigma}.
\ee
It is suggested that the reader shows that
\bea
\frac{\left<p\uparrow\right|\sum_{i=1}^{3}\mu_{iz}\left|p\uparrow\right>}{\left<n\uparrow\right|\sum_{i=1}^{3}\mu_{iz}\left|n\uparrow\right>} & = & -\frac{3}{2} \\
&{\mbox{\huge (}} \stackrel{\rm expt.}{=} & \frac{2.79}{-1.91}{\mbox{\huge )}}. \non
\eea

\subsubsection{Energy Levels}
Calculations in this model lead to expectations of finding the $L=1$, 
negative parity, states at $\hbar\omega$ and $N=1$ (1s) or $L=2$ (0d) 
states at $2\hbar\omega$.  Nature however presents us with two surprising results:
\begin{itemize}
  \item The {\em Roper}, P$_{11}$ (1450), which would naively be a 1s 
excitation, occurs {\bf below} the lowest negative parity 
``nucleon excited states'' -- the D$_{13}$ and S$_{11}$ at $\sim$1550 MeV.
\item There exists 200MeV between the 1s and 0d ``$2\hbar\omega$'' states 
N(1680) F$_{15}$.
\end{itemize}
In fact the Roper is still a mystery.  In the bag models it has been
described as a ``breathing mode'' \cite{Guichon:1985ny}, but it has also been
described as the result of coupling to the inelastic two-pion channels
\cite{Pearce:1986du,Schutz:1998jx}:
\bea
R & \rightarrow & N\pi\pi \non \\
& \rightarrow & N\eta \non
\eea
%


\section{One Gluon Exchange}

\begin{figure}[hbt]
\centering{\
	\epsfig{angle=0,figure=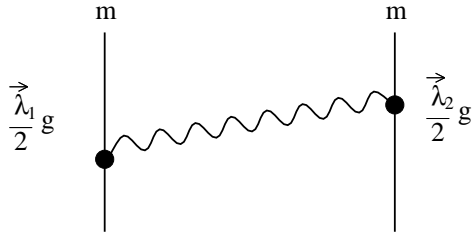, height=3cm} }
\parbox{100mm}{\caption{Picture of two quarks exchanging a gluon.}
\label{fig:gluon_exc}}
\end{figure}

We have so far considered only the simplest shell model picture of baryon
structure, with the quarks moving in a mean confining field. However,
one expects that there should be some residual interaction which, motivated 
by QCD, is usually taken to be the non-relativistic reduction of the One 
Gluon Exchange diagram shown in Fig. \ref{fig:gluon_exc}. This is 
proportional to the
product of two quark-gluon vertices, $\sum_a \vec{\lambda}_{1}^a
\vec{\lambda}_{2}^a = \vec{\lambda}_{1} \cdot \vec{\lambda}_{2}$. Using
the fact that the eigenvalue of the total colour wave function for a
baryon or meson must be zero, one easily finds:
\bea
\left<\vec{\lambda}_{1}\cdot\vec{\lambda}_{2}\right>_{\rm Baryons} & = &\half\left<\vec{\lambda}_{1}\cdot\vec{\lambda}_{2}\right>_{\rm Mesons} \non \\
& = & -\frac{8}{3}. \hspace*{5mm} \ex \\
\left({\rm Hint: }\sum_{a}\left(\lambda_{a}\right)^{2} = \frac{16}{3}\right)
\eea

Naturally this quantity is colour invariant, even though the $\vec{\lambda}$ 
are the colour matrices.  Defining the strong coupling 
constant in the usual way
\be
\alpha_{\rm s} = \frac{g^{2}}{4\pi},
\ee
we find that the one gluon exchange potential is given by
\bea
V_{\rm OGE}\left(\vec{r}\right) = -\frac{2}{3}\alpha_{\rm s}
\left[\frac{1}{r} - \frac{\pi}{m^{2}}\delta(\vec{r}) - 
\frac{1}{4m^{2}r^{3}}S_{12} - \frac{2\pi}{3m^{2}}\vec{\sigma}_{1}
\cdot\vec{\sigma}_{2}\delta(\vec{r})\right] \\ \non
- \frac{1}{4m^{2}r^{3}}\left(\left(\vec{\sigma}_{1}+
2\vec{\sigma}_{2}\right)\cdot\left(\vec{r}\times\vec{p}_{1}\right) - 
\left(\vec{\sigma}_{2} + 2\vec{\sigma}_{1}\right)\cdot
\left(\vec{r}\times\vec{p}_{2}\right)\right).
\eea
The terms inside the square brackets can be thought of as the Coulomb, 
Darwin and tensor ($S_{12} = (3\vec{\sigma}_{1}\cdot\hat{r}\vec{\sigma}_{2}
\cdot\hat{r}-\vec{\sigma}_{1}\cdot\vec{\sigma}_{2})$) terms 
respectively, while the final term ($\vec{\sigma}_{1}\cdot\vec{\sigma}_{2}$) 
is a hyperfine interaction.

\eg{}
{}For the $\Delta$, all $qq$ pairs have $S=1$, which implies that
\be
\left<\vec{\sigma}_{i}\cdot\vec{\sigma}_{j}\right>_{\Delta} = +1
\ee
For the $N$, there exists an equal probability of finding 
$S=1$ and $S=3$ pairs, so we find that
\be
\left<\vec{\sigma}_{i}\cdot\vec{\sigma}_{j}\right>_{N} = \half\left(+1-3\right) = -1
\ee
Therefore, when the hyperfine interaction is treated as a perturbation,  
$m_{\Delta}$ increases and $m_{N}$ decreases, by equal amounts.  
This effect breaks the degeneracy of the naive model.

One should note that on dimensional grounds
\be
\left<H_{\rm hyp}\right>_{(\mbox{$N$ or $\Delta$})} \propto \alpha_{S}\frac{\alpha^{3}}{m^{2}},
\ee
where the difference $m_{\Delta} - m_{N}$ determines the value 
of $\alpha_{s}$, which is typically of the order 0.6.

\eg{The $\Sigma$(1180)--$\Lambda$(1115) mass difference}
For the $\Sigma$ we have
\be
\begin{array}{c}
\left<\Sigma\right|(\vec{\sigma}_{u}\cdot\vec{\sigma}_{d}\left|\Sigma\right> = +1, \\
\left<\Sigma\right|\vec{\sigma}_{s}\cdot(\vec{\sigma}_{u}+\vec{\sigma}_{d}\left|\Sigma\right> = -4,
\end{array}
\ee
and therefore
\be
\left<H_{\rm hyp}\right>_{\Sigma} \propto \frac{\alpha_{s}\alpha^{3}}{\bar{m}^{2}} - \frac{4\alpha_{s}\alpha^{3}}{\bar{m}m_{s}}.
\ee

Similarly for the $\Lambda$ we find
\be
\begin{array}{c}
\left<\Sigma\right|(\vec{\sigma}_{u}\cdot\vec{\sigma}_{d}\left|\Sigma\right> = -3, \\
\left<\Sigma\right|\vec{\sigma}_{s}\cdot(\vec{\sigma}_{u}+\vec{\sigma}_{d}\left|\Sigma\right> = 0,
\end{array}
\ee
so
\be
\left<H_{\rm hyp}\right>_{\Lambda} \propto -\frac{3\alpha_{s}\alpha^{3}}{\bar{m}^{2}}.
\ee
Hence we see that $m_{\Sigma} > m_{\Lambda}$.


\section{Hadronic Shell Model}

A slightly more sophisticated approach is the hadronic shell model, 
suggested by Isgur and Karl\cite{Isgur:1978xj}.  Defining the 
Hamiltonian as
\be
H = \frac{\vec{p}_{\rho}{}^{2}}{2m} + \frac{\vec{p}_{\lambda}{}^{2}}{2m} + 
\sum_{i < j}\left[-\frac{2\alpha_{s}}{3r_{ij}} + \half br_{ij}\right] + 
H_{\rm hyp} + H_{\rm S_{12}}
\ee
with $\alpha_{s} \sim 0.6$,  we see that the $1/r_{ij}$ term is a colour
Coulomb interaction while $b$ represents a linear confining potential of 
strength $\sim 0.18$GeV$^{2}$.  The next step is to diagonalize 
$H$ in, say, a $2\hbar\omega$ space of harmonic oscillator wavefunctions.
This diagonalisation allows the calculation of baryon energy levels in
the model, however it is usually done with the caveat that the overall scale 
must be adjusted for each major shell.  
The detailed wave functions given by the model allow one to investigate
various electric and magnetic transition probabilities in detail.

\eg{ -- Neutron charge distribution}
As an exercise, the reader should, for a spin-up proton ($\uparrow$), show the probabilities of finding a particular quark to be $\uparrow$ or $\downarrow$ opposite a spin 1 or 0 pair are:
\be
\left.\begin{array}{ccc}
u_{1}^{\uparrow} = \frac{1}{18} &\hspace*{5mm} & u_{0}^{\uparrow} = \frac{1}{2} \\
u_{1}^{\downarrow} = \frac{2}{18} && u_{0}^{\downarrow} = 0 \\
d_{1}^{\uparrow} = \frac{2}{18} && d_{0}^{\uparrow} = d_{0}^{\downarrow} = 0 \\
d_{1}^{\downarrow} = \frac{4}{18} &&
\end{array}\right\}\ex
\ee
Similarly we find that for a spin-up, $\uparrow$, neutron
\be
\left.\begin{array}{ccc}
d_{1}^{\uparrow} = \frac{1}{18} &\hspace*{5mm} & d_{0}^{\uparrow} = \frac{1}{2} \\
d_{1}^{\downarrow} = \frac{2}{18} && d_{0}^{\downarrow} = u_{0}^{\uparrow} = u_{0}^{\downarrow} = 0 \\
u_{1}^{\uparrow} = \frac{2}{18} && \\
u_{1}^{\downarrow} = \frac{4}{18} &&
\end{array}\right\}\ex
\ee

For example, for the neutron, ($dd$) pairs are always $S=1$, (which feel a {\em repulsive} hyperfine interaction) while the ($ud$) pairs are 3:1, $(S=0):(S=1)$ (attractive : repulsive).
Therefore the effect of diagonalisation is to force 
the $dd$ pairs further apart.

\subsubsection{Open questions}

After studying these models one should be asking, among others, the following questions:
\begin{itemize}
  \item {\em What is the origin of $m$?}  At the 1GeV scale in QCD we have
\be
\begin{array}{rcl}
m_{u} &\sim& 6 {\rm MeV}, \\
m_{d} &\sim& 9 {\rm MeV}, \\
m_{s} &\sim& 150 {\rm MeV}.
\end{array}
\ee
  \item {\em What is the nature of confinement?}  For heavy quarks lattice
QCD suggests that:
\be
E(R) \sim -\frac{\alpha}{r} + cr,
\ee
captures the long and short distance pieces of the interaction.
On the other hand, phenomenologically the spacing between levels in
heavy quark systems ($b\bar{b}$ and $c\bar{c}$) 
is more or less independent of the quark mass, which
favours a fractional power dependence, like $r^{0.5}$ \cite{Motyka:1998rb}.

  \item {\em Is it really a few-body problem?}  This question is 
  directly related to the nature of the vacuum.
\end{itemize}

\section{Soliton Models}
\begin{figure}[hbt]
\centering{\
	\epsfig{angle=0,figure=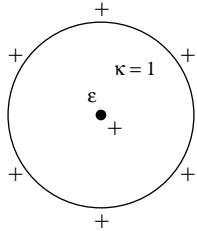, height=3cm} }
\parbox{100mm}{\caption{The dia-electric is anti-shielding.  Here $\kappa_{\rm med} \ll 1.$}
\label{fig:soliton_1}}
\end{figure}

One alternative to the models we have been discussing is the 
soliton model suggested by Lee \cite{Lee:1981mf}.  Suppose that 
the vacuum solution of QCD is a colour dia-electric ($\kappa < 1$). To
quote T. D. Lee:
\begin{quotation}
``Quark confinement is a large scale phenomenon.  Therefore, at least on phenomenological level, it should be understandable through a quasi-classical microscopic theory.''
\end{quotation}
To implement this concept, suppose that there exists a medium with 
$\kappa_{\rm med} \ll 1$.  In such a medium we find that any charge will 
produce a hole (vacuum) with $\kappa = 1$ (as illustrated in Fig. 
\ref{fig:soliton_1}).

It can be shown that the repulsion of $\varepsilon$ and the surface charges 
necessarily means that work has to be done to shrink the hole.
The continuity of the displacement, $\vec{D}$:
\be
\left.\vec{D}_{\rm in} (= \vec{E}_{\rm in}) = \vec{D}_{\rm out} (= \kappa_{\rm med}\vec{E}_{\rm out})\right|_{r=R},
\ee
means that the colour electric field is given by
\be
\begin{array}{rclc}
\vec{E}_{\rm in} &=& \frac{\varepsilon}{r^{2}}\hat{r} & \hspace*{5mm}r \leq R, \\
\vec{E}_{\rm out} &=& \frac{\varepsilon}{\kappa_{\rm med}r^{2}}\hat{r} & \hspace*{5mm}r \geq R.
\end{array}
\ee
Thus we find that
\bea
\left\{\mbox{Electric field energy of cavity}\right. & - & \left.\mbox{Electric energy without medium effect}\right\} \non \\
= \int_{\mbox{All Space}}\half\vec{E}\cdot\vec{D} & - & \left\{\mbox{Electric energy w/o medium effect}\right\}
\eea
Prove, as an exercise, that the above expression has the form
\be
\sim \varepsilon^{2}\frac{\left(\kappa_{\rm med}^{-1}-1\right)}{R} \longrightarrow \infty \hspace{3mm}{\rm as} \hspace{3mm} \kappa_{\rm med}\rightarrow 0, \ex
\ee
Hence we can see that a perfect dia-electric is {\em confining}.

\begin{figure}[hbt]
\centering{\
	\epsfig{angle=0,figure=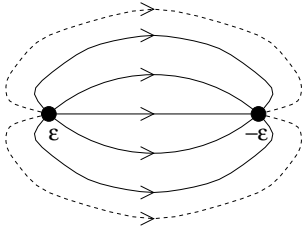, height=3cm} }
\parbox{100mm}{\caption{Solution form for the soliton model.  Here the dashed lines can be visualised as the self-consistent ``bag surface''}
\label{fig:soliton_2}}
\end{figure}

Now we suppose that the dia-electric vacuum is a {\em lower} energy state, so that it costs energy to make this hole.  The energy cost of making this hole is given by
\be
{\cal U}_{\rm hole} = BV + CS
\ee
where $V$ is the volume of the hole and $S$ is the surface energy.  Therefore we find the equilibrium radius, $R_{\rm equil}$, when
\be
\left.\frac{d}{dR}\left({\cal U}_{\rm hole} - {\cal U}_{\rm electric}\right)\right|_{R=R_{\rm equil}} = 0.
\ee
Naturally we have that $R_{\rm equil} \rightarrow \infty$ when 
$\kappa_{\rm med} \rightarrow 0$.  Therefore there are only solutions of 
the form illustrated by Fig. \ref{fig:soliton_2}, where the total
charge inside the cavity is zero.
This figure illustrates the close analogy between the superconducting
state, which excludes magnetic fields and the QCD vacuum which excludes
the colour electric field, leading to confinement.

\subsubsection{Practical Implementation}
These ideas were implemented in the model known as the Friedberg-Lee Soliton \cite{Lee:1981mf,Friedberg:1978sc}. 
\be
{\cal L}_{\rm FL} = \bar{q}(i\slsho{D}-m-f\sigma)q - 
\frac{1}{4}\kappa G_{a}^{\mu\nu}G^{a}_{\mu\nu} + 
\half(\del_{\mu}\sigma)^{2} - {\cal U}(\sigma)
\ee
where $\sigma$ is a new scalar field, and the term involving $\kappa$ is a dielectric function that we treat perturbatively for colourless states.
\begin{figure}[hbt]
\centering{\
	\epsfig{angle=0,figure=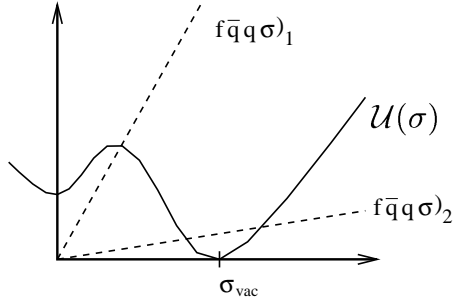, height=4cm} }
\parbox{100mm}{\caption{Illustration of the scalar field and
quark-scalar-field contributions to the total energy for two choices of
the valence quark, scalar density.}
\label{fig:soliton_3}}
\end{figure}

In Fig. \ref{fig:soliton_3} we see that the lowest energy state is 
when $q=0$, which is the non-perturbative vacuum with no valence quarks.

$\left.f\bar{q}q\right)_{1} \Rightarrow \sigma \approx 0$ which means that
when the scalar, valence quark density is high 
the perturbative vacuum is restored.

$\left.f\bar{q}q\right)_{2} \Rightarrow \sigma \approx \sigma_{\rm vac}$ 
which means that when the quark field is small the non-perturbative
vacuum is restored.

\begin{figure}[hbt]
\centering{\
	\epsfig{angle=0,figure=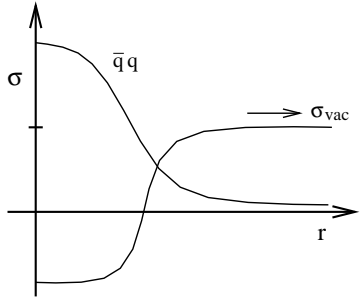, height=4cm} }
\parbox{100mm}{\caption{Formation of a non-topological soliton as a
self-consistent combination of localised valence quark density and a
``hole'' in the non-perturbative vacuum field configuration.}
\label{fig:soliton_4}}
\end{figure}

The results of this model are as follows
\begin{itemize} 
  \item The quarks dig a self-consistent ``hole'' in the scalar field.
  \item Asymptotically quarks have effective mass for $f\sigma_{\rm vac}$ ($+m$) (where $m$ is the perturbative QCD mass).
  \item Quarks are only ``confined'' in the sense that if, for example, $\kappa\sim(1-\frac{\sigma}{\sigma_{\rm vac}})^{n}$ then $\kappa \rightarrow 0$ outside the soliton.  Hence the total colour electric energy is infinite, 
unless $\sum_{i}\lambda_{i}^{a} = 0$ --- that is, it is a colourless state.  
This is shown by using the same argument as for a charge in a 
dia-electric cavity.
\end{itemize}


\section{Colour Dielectric Model}

Another interesting model is the Colour Dielectric Model\cite{Pirner:1984hd,Nielsen:1982fi,Dodd:1987pw}.This model gives us an effective description of the long-distance effects in QCD.  Once again, we can formulate a Lagrangian to allow the investigation of the theory:

\be
{\cal L} = i\bar{\psi}\slsh{\del}\psi - \frac{m}{\chi}\bar{\psi}\psi - U(\chi) + \half{\sigma_{\rm V}}^{2}\left(\del_{\mu}\chi\right)^{2} - \kappa(\chi){\rm Tr}(G_{\mu\nu}G^{\mu\nu}).
\ee

Here we find that $\sigma_{\rm V}\chi$ is a confining scalar field that mimics 
the non-perturbative effect of the gluons.  Note that when $\chi 
\rightarrow 1$ (inside the soliton) the second term goes to the usual
mass term, $m\bar{\psi}\psi$, while as $\chi \rightarrow 0$ the
effective mass of the quark goes to infinity, which certainly implies 
confinement:
\be
\chi \rightarrow 0 \hspace*{4mm} \Longrightarrow \hspace*{4mm} \frac{m}{\chi} \rightarrow \infty.
\ee
In the simplest case, the potential term, $U(\chi)$, may be taken to be 
quadratic in $\chi$: 
\be
U(\chi) \approx \half m_{\rm GB}^{2}{\sigma_{\rm V}}^{2}\chi^{2}
\ee

\section{MIT Bag Model}

The MIT bag model is of a similar vintage to the harmonic oscillator
shell model discussed earlier. We shall take some care in explaining it
because it is still proving to be extremely valuable in theoretical
nuclear physics. In particular, a large 
number of applications of the quark model of nucleon structure have been 
developed using the MIT bag.  Two that we mention especially are the 
calculation of nucleon properties in medium (for instance in a neutron 
star), and the Quark Meson Coupling (QMC) model \cite{Thomas:1984kv}.  
As a relativistic model which permits analytic solution, the MIT bag 
also allows us to examine the role of dynamical chiral symmetry breaking 
in hadron structure -- a topic of great current interest (for example)
under the heading chiral perturbation theory.
 
A model developed by Bogolubov \cite{Bogolubov:1968zk} in the late 1960's 
was the basis for the development of the MIT bag model.  The model was an
attempt to phenomenologically describe confined, relativistic quarks in a 
finite region of space.  We can view this model as, for example, an 
analytically solvable version of the Colour Dielectric model (where here 
$\chi$ is 0 outside and 1 inside the bag).  Bogolubov considered the 
simplest case: a Dirac particle of mass $m$, moving freely within a 
spherical volume of radius R, in a scalar potential.
\be
V_{S}(r) = -\theta(R-r)m
\ee
Inside the potential, the {\em effective} mass of the quark is 0, and outside the potential (or {\em bag}) it is infinite, thus confining the quark inside the bag.

Much of the structure of this model relies upon the time independent 
Dirac equation
\be
H\psi(\vec{r}) = E\psi(\vec{r}),
\label{eqn:mit_dirac}
\ee
where the Hamiltonian\footnote{See Appendix \ref{app:Math} for the mathematical conventions used in this paper.} is
\be
H = \vec{\alpha} \cdot \vec{p} + \beta (V_{S}(r) + m).
\ee

The eigenstates of the Hamiltonian are classified by finding operators that commute with $H$.  One of these operators is
\be
\vec{j} = \vec{l} + \frac{\vec{\sigma}_{(4\times4)}}{2},
\ee
with $\vec{\sigma}$ necessarily of the form
\be
\vec{\sigma}_{(4\times4)} = \left( \begin{array}{cc}
\vec{\sigma} & 0 \\
0 & \vec{\sigma}
\end{array} \right).
\ee
The other operator is the relativistic analog of an operator involving both 
spin and orbital angular momentum:
\be
k = \vec{\sigma} \cdot \vec{l} + 1.
\ee
The relativistic analog is the obvious generalisation
\bea
K & = & \gamma^{0}(\vec{\sigma} \cdot \vec{l} + 1) = \left( \begin{array}{cc}
k & 0 \\
0 & -k
\end{array} \right) \non \\
& = & \beta(\vec{\sigma} \cdot \vec{l} + 1).
\eea
So we can construct eigenstates of $\vec{l}+\vec{s}$ thus
\be
\left|l\:\half\:j\:\mu\right> \equiv \left|\chi_{\kappa}^{\mu}\right>,
\ee
and hence define
\be
K\left|\chi_{\kappa}^{\mu}\right> = -\kappa\left|\chi_{\kappa}^{\mu}\right>.
\ee
Using the above definitions, and the identities listed in the Appendix, it can be shown that $H$ does indeed commute with both $j$ and $K$, and that $j$ and $K$ commute with each other, that is
\be
\left[H,j\right] = 0 = \left[H,K\right] = \left[j,K\right]. \ex
\ee
The easiest way to obtain the eigenvalues of $K$ is to square it, and then 
operate on a wave function $\psi$. So we find:
\bea
K^{2} & = & \beta^{2}\left[(\vec{\sigma} \cdot \vec{l})^{2} + 2\vec{\sigma} \cdot \vec{l} + 1\right] \non \\
& = & l^{2} + \vec{\sigma} \cdot \vec{l} + 1 \non \\
& = & \vec{j}^{2} + \frac{1}{4},
\eea
and hence
\bea
K^{2}\psi & = & (\vec{j}^{2} + \frac{1}{4})\psi \non \\
& = & (j(j+1) + \frac{1}{4})\psi \non \\
& = & (j + \half)^{2}\psi.
\eea
We see that the eigenvalues of $K$ are $\kappa = \pm (j + \half)$, and thus
\be
\kappa = \left\{ \begin{array}{ll}
l    & 	{\rm for} \; j = l - \half \\
-l-1 &  {\rm for} \; j = l + \half
\end{array} \right.
\label{eqn:KappaVals}
\ee
thereby defining both $l$ and $j$ by the one quantum number $\kappa$.

Now, using
\be
\vec{\nabla} = \hat{r}\frac{\del}{\del r} - i \frac{\hat{r}}{r}\times\vec{l}, \ex
\ee
we can show that
\be
\vec{\alpha}\cdot\vec{p} = -i\vec{\alpha}\cdot\hat{r}\frac{\del}{\del r} + \frac{i}{r}\vec{\alpha}\cdot\hat{r}\left(\beta\kappa -1\right).
\ee
Substituting this back into Eq. (\ref{eqn:mit_dirac}) and solving we find that $\psi$ has a general solution of the form
\be
\psi_{\kappa}^{\mu} = \left[ \begin{array}{c}
g(r)\chi_{\kappa}^{\mu} \\
if(r)\chi_{-\kappa}^{\mu}
\end{array} \right],
\label{eqn:psiForm}
\ee
which satisfies the following coupled, ordinary differential equations
\bea
\left( E - V_{\rm s}(r) - m \right)g & = & -\left( \frac{df}{dr} + \frac{f}{r} \right) + \frac{\kappa f}{r}, \non \\
\left( E + V_{\rm s}(r) + m \right)f & = & \left( \frac{dg}{dr} + \frac{g}{r} \right) + \frac{\kappa g}{r}.
\label{eqn:ODE1}
\eea
For the most elementary case, $\kappa = -1$ (s$_{\half}$), these equations simplify to
\be
\frac{d^{2}u}{dr^{2}} + (E^{2} - (m + V_{\rm s})^{2})u = 0,
\ee
where the substitution $u=rg$ has been made.  Naturally there is a similar equation for $l$ $(=rf)$.  It is left as an exercise for the reader to show that by following the method of confinement, suggested by Bogolubov \cite{Bogolubov:1968zk}, which involves defining the scalar potential as
\be
V_{\rm s}(r) = \left\{ \begin{array}{rr}
{-m}    &  {\rm for} \; r \leq R \\
0       &  {\rm for} \; r > R,
\end{array} \right.
\label{eqn:W_eqn}
\ee
it can be show that, requiring $u$ and $l$ continuous at $r=R$ implies
\bea
\cos (\!ER) + \frac{\sqrt{(1-(E/m)^{2})}}{1+(E/m)}\sin (\!ER) = \frac{\sin (\!ER)}{ER}\left( 1 - \frac{E}{E+m}\right). 
\label{eqn:cont_eqn} \\
\ex \non
\eea
This gives us a boundary condition for this model.  By taking the 
confining limit ($m \rightarrow \infty$) we get the eigenvalue condition, 
given by the relationship between the spherical Bessel functions
\be
j_{0}(ER) = j_{1}(ER).
\ee
The energy levels may be parameterised by the definition
\be
E_{n\kappa} = \frac{\Omega_{n\kappa}}{R},
\ee
where $n$ is the principle quantum number.  We now find that the first two positive roots of the $\kappa = -1$ states are:
\bea
\Omega_{1-1} = 2.04 \;&\longleftrightarrow&\; 1\rm{s}_{\half},
\label{eqn:omegaVals} \\
\Omega_{2-1} = 5.40 \;&\longleftrightarrow&\; 2\rm{s}_{\half},
\eea
and further solutions are easily computed.

Thus Eq. (\ref{eqn:psiForm}) can be rewritten as
\bea
\psi_{n,\kappa}(\vec{r}) & \equiv & \psi_{n,-1}(\vec{r}) \non \\
& = & N_{n,-1}\left[ \begin{array}{c}
j_{0}\left(\frac{\Omega r}{R}\right) \\
i\vec{\sigma} \cdot \hat{r}j_{1}\left(\frac{\Omega r}{R}\right)
\end{array} \right] \chi^{\mu}_{-1}.
\label{eqn:psiSoln}
\eea

\begin{figure}[hbt]
\centering{\
	\epsfig{angle=0,figure=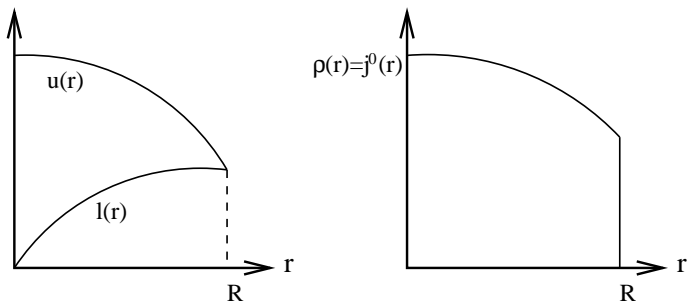, height=4cm} }
\parbox{100mm}{\caption{Pictorially we can see that $u(r) = l(r)$ at the boundary ($r=R$)}
\label{fig:bag_1}}
\end{figure}

Confinement is achieved in the MIT bag model by requiring that there is no quark current flow through the surface of the bag.  The (charge) density for the Dirac equation is given by
\bea
j^{0} & \equiv & \rho = \bar{\psi}\gamma^{0}\psi \non \\
& \propto & \left[ {j_{0}}^{2}\left(\frac{\omega r}{R}\right) + {j_{1}}^{2}\left(\frac{\omega r}{R}\right)\right],
\eea
so that 
the charge density is proportional to the baryon  number density.  
We also note that the space components of the current are given by
\be
\vec{j}(\vec{r}) = \bar{\psi}\vec{\gamma}\psi = \psi^{\dagger}\vec{\alpha}\psi,
\label{eqn:current_space}
\ee
so it is found that
\bea
\hat{r}\cdot\vec{j}(\vec{r}) & = & \left(j_{0}, -i\vec{\sigma}\cdot\hat{r}j_{1}\right)\left( \begin{array}{cc}
0 & \vec{\sigma}\cdot\hat{r} \\ 
\vec{\sigma}\cdot\hat{r} & 0
\end{array} \right)\left( \begin{array}{c}
j_{0} \\
i\vec{\sigma}\cdot\hat{r}j_{1}
\end{array} \right) \\
& = & ij_{0}j_{1} - ij_{0}j_{1} = 0.
\eea

In particular, at $r=R$, we can see that $\hat{r}\cdot\vec{j} = 0$, so there is no flow of quark current through the surface.

As a side issue, the reader might like to note that for a typical hadronic radius, $R\sim1$\,fm, the charge radius for a proton is
\be
\left<r^{2}\right>^{1/2}_{\rm ch} = 0.82{\rm fm},
\ee
and
\be
\frac{\Omega}{R} \sim 400{\rm MeV},
\ee
which should be compared with the constituent quark mass.

\subsubsection{Magnetic moment}

We now consider the case of a charged quark in this potential and introduce a constant magnetic field $\vec{B}$ to the system.  Naturally we have
\be
\vec{A} = \half\vec{B}\times\vec{r}. \ex
\label{eqn:A}
\ee
The quark current has a magnetic moment $\vec{\mu}$ that satisfies
\be
-\vec{\mu}\cdot\vec{B} = -\int dV \vec{j}\cdot\vec{A}. \non
\ee
By making a substitution for $\vec{j}$ (Eq. \ref{eqn:current_space}), and for $\vec{A}$ (Eq. \ref{eqn:A}) we have
\bea
-\vec{\mu}\cdot\vec{B} & = & -\frac{e}{2}\int dV \left(\psi^{\dagger}\vec{\alpha}\psi\right)\cdot\left(\vec{B}\times\vec{r}\right) \non \\
& = & - \frac{e}{2}\int dV \vec{r}\times \left(\psi^{\dagger}\vec{\alpha}\psi\right)\cdot\vec{B}.
\eea
Noting that both sides involve a dot product with $\vec{B}$ we find
\bea
\vec{\mu}& = & \frac{e}{2}\int dV \vec{r}\times\psi^{\dagger}\vec{\alpha}\psi \non \\
& & \vdots \hspace*{2mm}({\rm algebra}) \non \\
& = & \mu_{\rm conf}\vec{\sigma},
\eea
where
\bea
\mu_{\rm conf} & = & \frac{eR}{2\Omega} \cdot \left\{\frac{4\Omega - 3}{6(\Omega - 1)}\right\} \\
& = & \frac{e}{2``m_{\rm const}{\char 34}} \left\{0.83\right\},
\eea
and as noted earlier, $m_{\rm const} \equiv \Omega/R$ is like a
constituent quark mass.
One should note that the previous relationship for 
$\frac{\mu_{p}}{\mu_{n}} (=-\frac{3}{2})$ is preserved.

\subsubsection{Axial current}

We know that axial current is very important for weak interactions, and that it is calculated by
\be
{\bf A}^{\mu} = \bar{\psi}\gamma^{\mu}\gamma_{5}\frac{{\mbox{\boldmath $\tau$}}}{2}\psi,
\ee
where \mbox{\boldmath $\tau$} is the 3 Pauli matrices for isospin and
\be
\psi = \left(\begin{array}{c}
u \\
d
\end{array}\right),
\ee
where both $u$ and $d$ are a four component spinor.

In the non-relativistic limit $\vec{A}_{i}$ dominates.  The gamma matrices give us
\be
\gamma^{0}\vec{\gamma}\gamma_{5} = \left(\begin{array}{cc}
1 & 0 \\
0 & -1
\end{array}\right)\left(\begin{array}{cc}
0 & \vec{\sigma} \\
-\vec{\sigma} & 0
\end{array}\right)\left(\begin{array}{cc}
0 & 1 \\
1 & 0
\end{array}\right) = \left(\begin{array}{cc}
\vec{\sigma} & 0 \\
0 & \vec{\sigma}
\end{array}\right),
\ee
and in the non-relativistic situation we have $\psi$ nearly, but not exactly equal to $\left(\begin{array}{c} 1 \\ 0\end{array}\right)$, so
\be
\vec{A}_{i} \simeq \vec{\sigma}_{i}\frac{\tau_{i}}{2}
\ee
with equality when $\psi = \left(\begin{array}{c} 1 \\ 0\end{array}\right)$ -- 
the index $i$ is an index in isospin.

Another exercise for the reader is to show, for the non-relativistic quark 
model, using the spin-isospin wave function obtained earlier, that:
\be
\frac{\left<p\uparrow\right|\sum_{j=1}^{3}\frac{\sigma_{jz}\tau_{j3}}{2}\left|p\uparrow\right>}{\left<p\uparrow\right|\frac{\sigma_{z}\tau_{3}}{2}\left|p\uparrow\right>} = \frac{5}{3}, \ex
\ee
where the index $j$ labels the quarks, and the denominator involves simply the baryon spin-isospin operators.
Thus, with the spin-flavor wave functions we have used and the nucleon axial charge operator, we find that
\be
\frac{g_{\rm A}}{g_{\rm V}} = 
\left(\frac{\left<p\uparrow\right|\vec{{\bf A}}\left|p\uparrow\right>}{\left<p\uparrow\right|\frac{\vec{\sigma}{\mbox{\boldmath $\tau$}}}{2}\left|p\uparrow\right>}\right) = \frac{5}{3}.
\ee
However, from the $\beta$-decay of free neutrons we know that it is 1.26!

One suggestion to remedy this problem with $g_A$ was that there might be
a strong tensor force which would mix a sizeable d-state (L=2) component
into the nucleon wavefunction, thus reducing $g_A$. Exploration of the
deformation of the $\Delta$, through the E2/M1 ratio in the $\Delta
\rightarrow N \gamma$ transition has since suggested that this is
unlikely to be the correct explanation -- the N and $\Delta$ do not
appear to have a significant, intrinsic deformation.

On the other hand, for our case the quarks are {\bf not} non-relativistic:
\bea
\bar{\psi}\vec{\gamma}\gamma_{5}\frac{\bm{\tau}}{2}\psi & = & \left(j_{0}, -i\vec{\sigma}\cdot\hat{r}j_{1}\right)\left( \begin{array}{cc}
\vec{\sigma} & 0 \\ 
0 & \vec{\sigma}
\end{array} \right)\left( \begin{array}{c}
j_{0} \\
i\vec{\sigma}\cdot\hat{r}j_{1}
\end{array} \right) \\
& = & \left[\left\{j_{0}\left(\frac{\Omega r}{R}\right)\right\}^{2}\vec{\sigma} + (\vec{\sigma}\cdot\hat{r})\vec{\sigma}(\vec{\sigma}\cdot\hat{r})\left\{j_{1}\left(\frac{\Omega r}{R}\right)\right\}^{2}\right]\frac{\bm{\tau}}{2},
\eea
and
\bea
\int dV \bar{\psi}(\vec{r})\vec{\gamma}\gamma_{5}\frac{\bm{\tau}}{2}\psi(\vec{r}) & = & {\cal N}^{2}\int_{0}^{R}dr r^{2}\left[\left\{j_{0}\left(\frac{\Omega r}{R}\right)\right\}^{2}\right. \non \\
& & \left.- \frac{1}{3}\left\{j_{1}\left(\frac{\Omega r}{R}\right)\right\}^{2}\right]\vec{\sigma}\frac{\bm{\tau}}{2}, \\
& & \ex \non
\eea
(For reference ${\cal N}^{2} = \left[\int_{0}^{R}dr r^{2}\left({j_{0}}^{2} + {j_{1}}^{2}\right)\right]^{-1} )$
\bea
\stackrel{\rm (numerical)}{=}  0.65 \vec{\sigma}\frac{\bm{\tau}}{2}. \non
\eea
This implies that
\bea
\frac{g_{\rm A}^{\rm bag}}{g_{\rm V}} & = & 0.65 \times \frac{5}{3} \non \\
& = & 1.09,
\eea
a result which constitutes nearly a 50\% improvement in the error and thus may be
considered a major success of relativistic quantum mechanics.  
However this success is tempered with the knowledge that the quarks were 
taken to be massless.
If we were to give the quarks a ``current quark'' mass, the lower component 
would decrease and hence $g_{\rm A}$ increase, so that
in the heavy quark limit $g_{\rm A}$ increases to $\frac{5}{3}$.

This model of Bogolubov produces many interesting results, but the value of the bag radius, $R$, is chosen in an ad hoc manner.  What is needed is a self-consistent link between the size of the cavity, and what is in it, similar to the soliton models.  This self-consistency was supplied by the MIT bag model in a fully
covariant way.  We specialise to the spherical, static cavity (which is the 
only case that is easily solved in 3+1 dimensions
\footnote{The deformed case has been looked at by Viollier, Kerman, and
others} ).

\subsubsection{Lagrangian formulation}

In the static cavity approximation the Lagrangian density describing the
MIT bag is:
\be
{\cal L}_{\rm cav} = \left[\bar{q}(i\slsh{\del}-m)q - B\right]\theta_{\rm V} - \half \bar{q}q\delta_{\rm S},
\ee
where the last term results in the quarks being infinitely massive at the 
surface of the cavity.  For the present we have not 
included the gluons for simplicity.
(To include the gluons we must 
include a term $-\frac{\kappa}{4}{\rm Tr}(G\cdot G)$, 
with 
\be
\kappa = \left\{ \begin{array}{ll}
1 & \;\mbox{ inside the cavity} \\
0 & \;\mbox{ outside the cavity}
\end{array} \right.
\ee
and make the substitution $\del \rightarrow D$.)
We have defined:
\bea
\theta_{\rm V} & = & \theta(R-r), \\
\delta_{\rm S} & = & \delta(r-R),
\eea
and the Euler-Lagrange equation
\be
\frac{\del{\cal L}}{\del\bar{q}} - \del_{\mu}\frac{\del{\cal L}}{\del(\del_{\mu}\bar{q})} = 0,
\ee
is satisfied.  We demand that the action is stationary under the following transformations
\bea
q & \rightarrow & q + \delta q, \non \\
\bar{q} & \rightarrow & \bar{q} + \delta\bar{q}, \non \\
R & \rightarrow & R + \delta R.
\eea
The final transformation results in
\bea
\theta_{\rm V} & \rightarrow & \theta_{\rm V} + \delta_{\rm S}(\delta R), \non \\
\delta_{\rm S} & \rightarrow & \delta_{\rm S} - n\cdot\del\delta_{\rm S},
\eea
where $n^{\mu} = (0,\hat{r})$ is defined to be an outward normal.

{\bf Exercise \mbox{\exv}} for the reader is to show that these transformations lead to the following equations
\be
\left.\begin{array}{cl}
i\slsh{\del}q = mq  & r \leq R, \\
& \mbox{Dirac Equation}\\
i\gamma \cdot n q = q & r = R, \\
&\mbox{linear boundary condition (l.b.c.)} \\
B = -\half n\cdot\del(\bar{q}q), & r = R, \\
& \mbox{non-linear boundary condition (n.l.b.c.)}
\end{array}\right\}\bm{{\rm Ex \,\char 35}\arabic{exercise}}\addtocounter{exercise}{1}
\ee

{\bf Exercise \mbox{\exv}} is to show that the linear boundary condition implies that at $r=R$,
\bea
n_{\mu}j^{\mu} & = & +\bar{q}q \non \\
& = & -\bar{q}{q} \non \\
& = & 0, \ex
\eea

As $\bar{q}q$ vanishes at $r=R$, it can be shown that the total energy of 
three identical quarks is given by
\be
E(R) = 3\frac{\Omega}{R}+\frac{4\pi}{3}R^{3}B,
\ee
where the first term comes from the Dirac equation and the second term can be thought of as the volume energy.  We also find the non-linear boundary condition is the same as the Lee soliton, that is
\bea
\frac{\del E(R)}{\del R} = 0& \Rightarrow & \frac{3\Omega}{R^{2}} = 4\pi BR^{2} \non \\
& \Rightarrow & R = \left(\frac{3\Omega}{4\pi B}\right)^{1/4}
\eea
Thus we can choose $B$ once and fix it, and then calculate the size of each of the hadrons since the radius is determined by what energy state the quarks are in.  To get a instinctive feeling for this equation one can visualise it as requiring that the pressure from the vacuum is balanced by the energy of the quarks at the surface of the bag.

Since this model, as described, is quite simplistic, possible improvements are easy to imagine and often easy to implement.  One of the most important extensions was the inclusion of the zero point and centre of mass corrections
\be
E(R) \rightarrow E(R) - \frac{z_{0}}{R}.
\ee
Also one gluon exchange may be included by solving Maxwell's equation in a 
cavity, subject to the confining boundary conditions.  This leads to a
hyperfine interaction  which has the same dependence on spin and colour
as the hyperfine interaction discussed earlier and hence we get the 
required $N-\Delta$, $\Sigma-\Lambda$ mass splitting, etc.  

Given ${\cal L}_{\rm cav}$ we can follow the common procedure used in QCD 
and look at the symmetries, and hence the conserved currents.  
Recall Noether's theorem, that if ${\cal L}(\psi_{i},\del_{\mu}\phi_{i})$ 
is invariant (i.e. $\delta{\cal L}=0$) under the transformation
\be
\phi_{i} \rightarrow \phi_{i} + f_{i}(\phi_{j})\varepsilon,
\ee
where $\varepsilon$ is an infinitesimal constant, then
\be
j^{\mu} = \frac{\del{\cal L}}{\del(\del_{\mu}\phi_{i})}f_{i}
\ee
is conserved.  Thus we have that if ${\cal L} = {\cal L}_{\rm symm} + {\cal L}_{\rm break}(\phi_{i})$, then from the Euler-Lagrange equation:
\be
\del_{\mu}j^{\mu} = \frac{\del{\cal L}_{\rm break}}{\del\phi_{j}}f_{j}.
\ee

We now will look at some of the symmetries in the model.
\eg{U(1) invariance}
It is easy to show that the Lagrangian density
\be
{\cal L}_{\rm cav} = \left[i\bar{q}\slsh{\del}q - B\right]\theta_{V} - \half\bar{q}q\delta_{\rm S},
\ee
is invariant under the phase transformations
\bea
q & \rightarrow & q + i\varepsilon q, \non \\
\bar{q} & \rightarrow & \bar{q} - i\varepsilon \bar{q}.
\eea
Thus we find that the baryon number current (up to a factor $1/3$): 
\bea
j^{\mu} & = & \left(i\bar{q}\gamma^{\mu}\right)(-iq)\theta_{\rm V} \non \\
& = & \bar{q}\gamma^{\mu}q\theta_{\rm V}, 
\eea
is conserved.

\eg{ -- Isospin Current}
If we define $q = \left(\begin{array}{c}u \\ d\end{array}\right)$, then we find that ${\cal L}_{\rm cav}$ is invariant under
\bea
q & \rightarrow & q + i\frac{\bm{\tau}\cdot\bm{\varepsilon}}{2}q, \non \\ 
\bar{q} & \rightarrow & \bar{q} - 
i\bar{q}\frac{\bm{\tau}\cdot\bm{\varepsilon}}{2}.
\eea
It is left as an exercise for the reader to check that the 
isospin current is conserved and is given by
\be
j^{\mu} = \bar{q}\gamma^{\mu}\frac{\bm{\tau}}{2}q\theta_{\rm V}. \ex
\ee

\begin{figure}[hbt]
\centering{\
	\epsfig{angle=0,figure=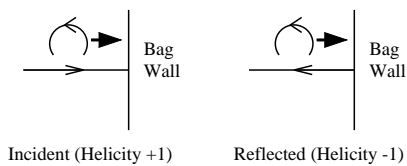, height=2cm} }
\parbox{100mm}{\caption{Violation of chiral symmetry at the bag surface}
\label{fig:a}}
\end{figure}

\eg{ -- Axial Current}
For two-flavour, massless, QCD, ${\cal L}$ is invariant under
\bea
q & \rightarrow & q - i\frac{\bm{\tau}\cdot\bm{\varepsilon}}{2}\gamma_{5}q, \non \\
\bar{q} & \rightarrow & \bar{q} - i\bar{q}\frac{\bm{\tau}\cdot\bm{\varepsilon}}{2}\gamma_{5},
\label{eqn:chiral_qcd}
\eea
which are chiral transformations, and there exists a conserved axial current
\be
\bm{A}^{\mu} = \bar{q}\gamma^{\mu}\gamma_{5}\frac{\bm{\tau}}{2}q.
\ee
However for ${\cal L}$ the surface term ``$-\half\bar{q}q\delta_{\rm S}$'' is not invariant under the chiral transformation, Eq. (\ref{eqn:chiral_qcd}), and therefore
\be
\bm{A}^{\mu} = \bar{q}\gamma^{\mu}\gamma_{5}\frac{\bm{\tau}}{2}q\theta_{\rm V},
\ee
satisfies
\be
\del_{\mu}\bm{A}^{\mu} = -i\bar{q}\gamma_{5}\frac{\bm{\tau}}{2}q\delta_{\rm S},
\ee
and we see (as illustrated in Fig. \ref{fig:a}) that the confining boundary 
violates chiral symmetry -- i.e. mixes left- and right-handed quarks.
This is a major problem since symmetries should always be a crucial
guide in constructing models.

But QCD already has a problem.  If $\del_{\mu}\bm{A}^{\mu} = 0$ then
\be
\int_{\mbox{All Space}}dV\del_{\mu}\bm{A}^{\mu} = 0,
\ee
implies
\be
\del_{0}\left\{\int dV \bm{A}^{0}\right\} = -\int dV \vec{\nabla}\cdot\vec{\bm{A}} = 0,
\ee
by the Gauss Theorem.  By defining the axial charge as $\bm{Q}_{5} = \int dV \bm{A}^{0}$ we find that it is a constant of the motion, that is
\be
\left[H,\bm{Q}_{5}\right] = 0.
\ee
Thus for all positive parity eigenstates of the Hamiltonian there exists a degenerate, negative parity state, i.e.
\bea
H\left|N^{+}\right> = m\left|N^{+}\right> \Rightarrow \bm{Q}_{5}\left|N^{+}\right> & = & \left|N^{-}\right> \; {\rm has} \non \\
H\left|N^{-}\right> & = & m\left|N^{-}\right>.
\eea
This is clearly not seen in nature!

The solution comes through the Goldstone Theorem. 
Either these degenerate, negative parity states exist 
{\bf OR} $\bm{Q}^{5}\left|0\right> \neq 0$.  That is, as a result of 
spontaneous symmetry breaking, there exists massless, pseudoscalar
Goldstone bosons. As the first option is clearly incorrect, 
in an effective low energy description of
QCD, we require a 
massless ``pion'' field, $\bm{\phi}$, in addition to the quarks. 
so that ${\cal L}_{\rm cav}$ is invariant.  Naturally, if we probe this at 
high momentum transfer the $\pi$ will show its internal $\bar{q}q$ structure. 
This is the basis of the Cloudy Bag Model \cite{Theberge:1980st}
discussed in Sec. \ref{sec:CBM}, 
which has a Lagrangian of the form
\be
{\cal L}_{\rm CBM} = \left[i\bar{q}\slsh{\del}q - B\right]\theta_{\rm V} - \half\bar{q}e^{i\bm{\tau}\cdot\bm{\phi}\gamma_{5}/f}q\delta_{\rm S} + \half\left(D_{\mu}\bm{\phi}\right)^{2},
\ee
where
\bea
D_{\mu}\bm{\phi} & = & \left(\del_{\mu}\phi\right)\hat{\phi} + f\sin\left(\frac{\phi}{f}\right)\del_{\mu}\hat{\phi}, \\
& = & \del_{\mu}\bm{\phi} + {\cal O}(\phi^{3}).
\eea
and ${\cal L}_{\rm CBM}$ is invariant under
\bea
q & \rightarrow & q - i\frac{\bm{\tau}\cdot\bm{\varepsilon}}{2}\gamma_{5}q, \non \\
\bar{q} & \rightarrow & \bar{q} - i\bar{q}\frac{\bm{\tau}\cdot\bm{\varepsilon}}{2}\gamma_{5}, \non \\
\bm{\phi} & \rightarrow & \bm{\phi} + \bm{\varepsilon}f + f\left(\bm{\varepsilon}\times\hat{\phi}\right)\times\hat{\phi}\left[1-\frac{\phi}{f}\cot(\frac{\phi}{f})\right].
\eea
Thus
\bea
\bm{A}^{\mu} & = & \bar{q}\gamma^{\mu}\gamma_{5}\frac{\bm{\tau}}{2}q\theta_{\rm V} + \left[f\hat{\phi}(\del^{\mu}\phi) + \frac{f^{2}}{2}\del^{\mu}\hat{\phi}\sin(\frac{2\phi}{f})\right] \non \\
& = & \bm{A}^{\mu}_{\rm quark} + f\del^{\mu}\bm{\phi} + {\cal O}(\phi^{3}).
\eea

\section{The Cloudy Bag Model (CBM)}
\label{sec:CBM}
The basic premise for this model is that we work perturbatively about 
MIT bag model solutions.  This can be compared to the Skyrme models 
or the topological soliton models, where the pion field dominates the 
dynamics.  The question remains as to whether they are equivalent.  For QCD 
in 1+1 dimensions this equivalence can be demonstrated, but in 3+1 
dimensions the belief is that the equivalence is broken.

The CBM Lagrangian is defined to be
\bea
{\cal L}_{\rm CBM} & \cong & \left(i\bar{q}\slsh{\del}q - m\bar{q}q - B\right)\theta_{\rm V} - \half\bar{q}q\delta_{\rm S} - \non \\
& &\frac{i}{2f}\bar{q}\gamma_{5}\frac{\bm{\tau}}{2}q\cdot\bm{\phi}\delta_{\rm S} + \half\left(\del_{\mu}\bm{\phi}\right)^{2} - \half m_{\pi}^{2}\bm{\phi}^{2},
\eea
and here the terms involving $m$ use the current quark mass to 
slightly (partially) break exact chiral symmetry.  
We also see that chiral symmetry requires (at least) linear coupling of pions 
to the bag through the $\bm{\tau}\cdot\bm{\phi}$ term.  That is, the 
CBM Lagrangian reduces to the MIT bag plus a free pion field with
interactions between the two dictated by chiral symmetry.
\be
{\cal L}_{\rm CBM} \cong {\cal L}_{\mbox{MIT Bag}} + {\cal L}_{\rm Free-\pi} + {\cal L}_{\rm int}.
\ee
Hence this implies that
\bea
H & = & H_{\rm MIT} + H_{\pi} + H_{\rm int} \\
& = & \sum_{\alpha} \varepsilon_{\alpha}\alpha^{\dagger}\alpha+\sum_{k}\omega_{k}a_{k}^{\dagger}a_{k}+\sum_{\alpha,\beta,k}\left[\beta^{\dagger}\alpha a_{k}v^{\beta\alpha}_{k} + {\rm h.c.}\right],
\eea
where ($k=\vec{k},i$) and
\be
v^{\beta\alpha}_{\vec{k},i} = \frac{i}{2f}\frac{1}{\sqrt{2\omega_{k}}}\int dV \exp^{-i\vec{k}\cdot\vec{r}}\delta(r-R)\left<\beta\right|\bar{q}(\vec{r})\gamma_{5}\bm{\tau}_{i}q(\vec{r})\left|\alpha\right>,
\ee
with $\left<\beta\right|$ and $\left|\alpha\right>$ being bag states.
We stress that the pion-baryon couplings involve form factors which
arise naturally as a consequence of the internal structure of the
baryons and which can be calculated using the underlying quark model --
be it a bag or something more sophisticated.

Another exercise for the reader is, for $\alpha=\left|N\right>$, and 
$\beta=\left|N^{\prime}\right>$, using
\be
N^{2}_{n,\kappa} \equiv N^{2}_{1,-1} = \frac{\Omega^{3}}{2R^{3}(\Omega -1)\sin^{2}\Omega},
\ee
to prove that 
\bea
v_{\vec{k},i}^{N^{\prime}N} & = & -\left(2\omega_{k}\right)^{-1/2}\frac{i}{2f}\frac{\Omega}{(\Omega-1)}\frac{j_{1}(kR)}{kR}  \non \\
& &{}_{SI}\left<N^{\prime}\right|\sum_{\alpha =1}^{3}\tau_{\alpha i}\vec{\sigma}_{\alpha}\cdot\vec{k}\left|N\right>_{\rm SI}.\ex
\eea
Hence we see that using $g_{\rm A}^{\rm Bag} = \left(\frac{\Omega}{\Omega -1}\right)\frac{5}{9}$, and
\be
u(k) \equiv \frac{3j_{1}(kR)}{kR} \;\left(\stackrel{k\rightarrow 0}{\longrightarrow}1\right),
\ee
it can be shown
\be
v_{\vec{k},i}^{N^{\prime}N}=-\frac{i}{(2\omega_{k})^{1/2}}\left(\frac{g_{\rm A}^{\rm Bag}}{2f}\right)u(k)\left<N^{\prime}\right|\tau_{i}\vec{\sigma}\cdot\vec{k}\left|N\right>,
\ee
and comparing to the usual $\pi NN$ coupling, we find 
\be
\frac{g_{\pi NN}}{2m_{N}} = \frac{g_{\rm A}}{2f}.
\ee
With a little rearrangement,
we find that
\be
fg_{\pi NN} = g_{\rm A}m_N,
\ee
which is the Goldberger-Treiman relation.

{}From this same ${\cal L}$ we can derive various coupling constants and form-factors, for instance, $\Delta N\pi, \Delta\Delta\pi, \Sigma\Lambda\pi, \Sigma\Sigma\pi, \Xi\Xi\pi,$ etc.

\eg{The physical/dressed nucleon}
We have included some examples that may assist in the understanding of 
this model here. The physical nucleon, $\left|\tilde{N}\right>$, is the
eigenstate of the CBM Hamiltonian with eigenvalue $m_N$:
\be
H\left|\tilde{N}\right> = m_{\rm N}\left|\tilde{N}\right>.
\ee
This should be compared with the eigenstate of the MIT bag piece of the
Hamiltonian:
\bea
H_{\rm MIT}\left|N\right> = m_{\rm N}^{(0)}\left|N\right>, \non \\
H_{\rm MIT}\left|\Delta\right> = m_{\Delta}^{(0)}\left|\Delta\right>, \non \\
{\rm e.t.c.}
\eea
where the one gluon exchange splittings, which may or may not be equal
to the mass splittings of the observed baryons, 
is included in the $m^{(0)}$ terms.  In first order perturbation theory:
\bea
\left|\tilde{N}\right> & = & \sqrt{Z}\left\{\left|N\right> + 
\sum_{\alpha,k}\frac{\left<\alpha\pi_{k}\right|H_{\rm int}\left|N
\right>}{m_{\rm N}-w_{k}-m_{\alpha}}\left|\alpha\pi_{k}\right>\right\} \\
& \cong & \sqrt{Z}\left\{\left|N\right> + \sum_{k}c_{1}(k)\frac{\left|
N\pi_{k}\right>}{-\omega_{k}} + \sum_{k}c_{2}(k)\frac{\left|\Delta\pi_{k}
\right>}{m_{\rm N}-\-m_{\Delta}-\omega_{k}}\right\}.
\eea
The convergence of this perturbative expansion can be proven to be
extremely good as long as $R$ is large ($\geq 0.7$fm) -- i.e. 
the form-factor is soft -- and as long as we restrict our model space to 
low-lying bag states (e.g. $N, \Delta[R,\ldots]$) -- similar to 
the $2\hbar\omega$ restriction in the non-relativistic oscillator model.

\eg{ -- Charge distribution of $n$}
Suppose we ignore spin-spin effects (small in the bag), then we find that 
the neutron bag has no charge distribution. On the other hand, the wave
function of the physical neutron is:
\be
\left|\tilde{n}\right> = \sqrt{z}\left\{\left|n\right> + \sum_{k}c_{1}(k)\left[\left(\frac{2}{3}\right)^{1/2}\left|p\pi_{k}^{-}\right> + \left(\frac{1}{3}\right)^{1/2}\left|n\pi^{0}\right>\right] + \ldots\right\}.
\ee
Here we see that $\left|n\right>$ contributes nothing as does 
$\left|n\pi^{0}\right>$.  The coefficients $\frac{2}{3}$ and $\frac{1}{3}$ 
should be instantly recognizable as the Clebsch-Gordon coefficients for 
isospin coupling.  Since there is only one term ($\left|p\pi^{-}_{k}\right>$) 
that contributes (at this order) we can now see that the {\bf leading 
order effect} in producing a non-zero charge distribution for the 
neutron is the (proton bag) + (pion cloud) component of the neutron 
wave function \cite{Theberge:1982st}.

\begin{figure}[hbt]
\centering{\
\epsfig{angle=0,figure=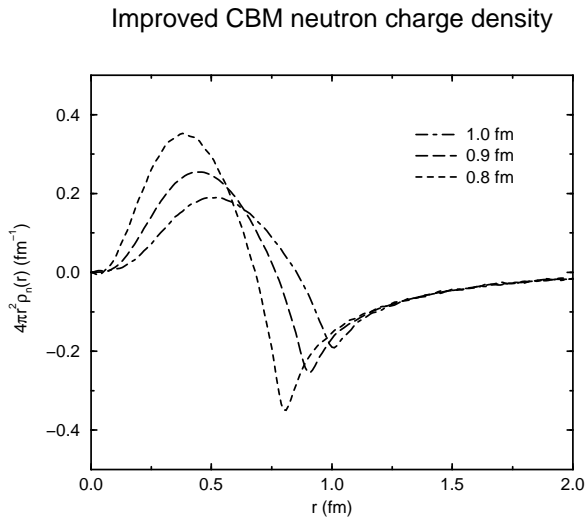, height=7cm} }
\parbox{100mm}{\caption{Neutron charge density calculated in the CBM for
several choices of bag radius. Note that the peak in the negative charge
density always occurs very near the chosen value of R.}
\label{nchff}}
\end{figure}

It is easy to see that, as the bag has a sharp surface, this simple
model will produce a zero in the neutron charge density at the bag
radius \cite{Theberge:1982st}. 
Detailed calculations of the neutron charge distribution are
complicated by the need to project a state of good total momentum and to
remove spurious centre of mass motion (which cannot be treated exactly
as it was in the oscillator model). Nevertheless, after all the
necessary corrections are made this feature persists to quite good
accuracy. In Fig.~\ref{nchff}, we see the results of a very recent
calculation \cite{Lu:1997sd}, which illustrate the importance of the current
measurements of the neutron charge form factor at MIT Bates, TJNAF and
Mainz.

\eg{ -- Corrections to bag mass}
To lowest, non-trivial, order (second order in $H_{\rm int}$) we have 
\bea
m_{N} & = & m_{N}^{(0)} + \frac{3f_{NN\pi}^{2}}{\pi m_{\pi}^{2}}\int_{0}^{\infty}dk\frac{k^{4}u^{2}(k)}{\omega_{k}(m_{N}-\omega_{K}-E_{N}(k))} \non \\
& & + \frac{4}{3}\frac{f_{N\Delta\pi}^{2}}{\pi m_{\pi}^{2}}\int_{0}^{\infty}dk\frac{k^{4}u^{2}(k)}{\omega_{k}(m_{N}-\omega_{K}-E_{\Delta}(k))}.
\eea

\begin{figure}[hbt]
\centering{\
	\epsfig{angle=0,figure=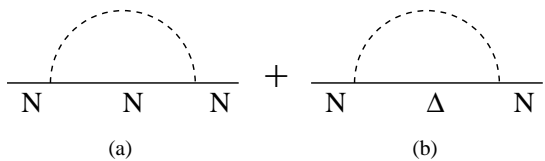, height=2cm} }
\parbox{100mm}{\caption{Pion self-energy corrections for the nucleon}
\label{fig:bag1}}
\end{figure}

Since both these diagrams are attractive there is a sizable correction of -300 to -400 MeV to the total energy.  

For comparison we look at the $\Delta-\Delta$ system.
\begin{figure}[hbt]
\centering{\
	\epsfig{angle=0,figure=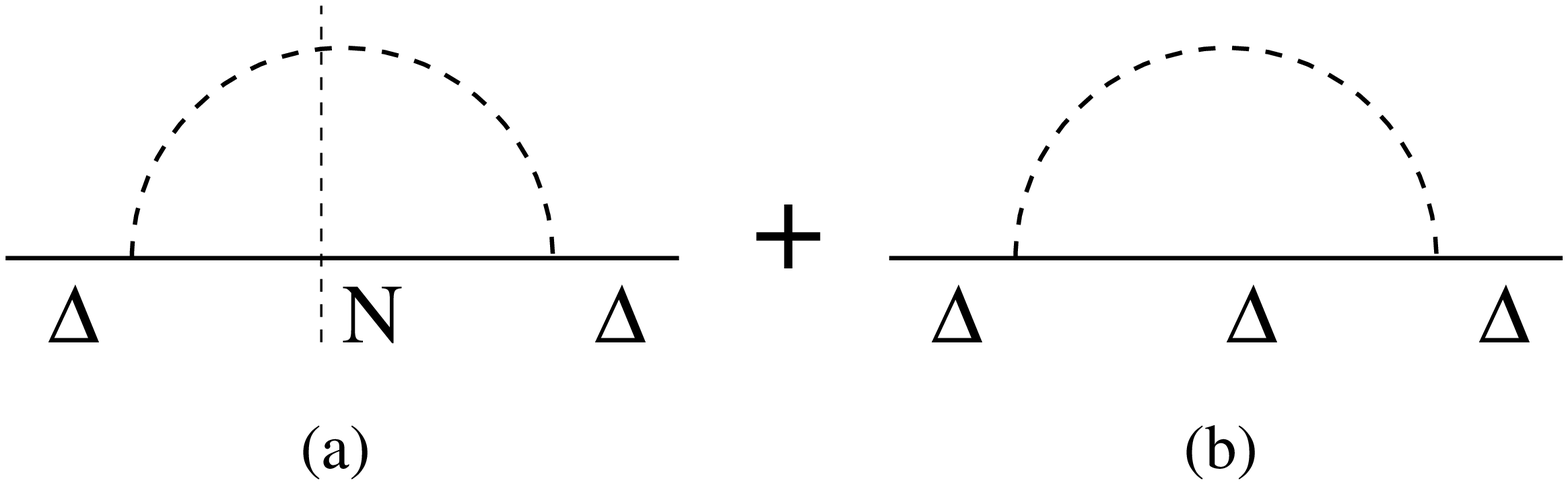, height=2cm} }
\parbox{100mm}{\caption{Pion self-energy corrections for the $\Delta$}
\label{fig:bag2}}
\end{figure}
It can be show that Fig. \ref{fig:bag1}(a) and Fig. \ref{fig:bag2}(b) are equal in magnitude and that Fig. \ref{fig:bag2}(a) is small.  So we find that
\bea
{\rm Re}(m_{\Delta}) & = & m_{\Delta}^{(0)} + \frac{f_{\Delta N\pi}^{2}}{3\pi m_{\pi}^{2}}{\cal P}\left[\int_{0}^{\infty}dk\frac{k^{4}u^{2}(k)}{\omega_{k}(m_{\Delta}-\omega_{k}-E_{N}(k))}\right] \non \\
& & + \frac{75}{16}\frac{f_{\Delta\Delta\pi}^{2}}{\pi m_{\pi}^{2}}{\cal P}\left[\int_{0}^{\infty}dk\frac{k^{4}u^{2}(k)}{\omega_{k}(m_{\Delta}-\omega_{k}-E_{\Delta}(k))}\right],
\eea
where ${\cal P}$ is the Principal value of the integral and in this case results in severe cancellations.  The first integral term is represented by Fig. \ref{fig:bag2}(a) and the second integral represents Fig. \ref{fig:bag2}(b).  In this case the attraction leads to a downward shift of 200 to 300 MeV!

Hence we can see that 100 to 200 MeV of the $N-\Delta$ splitting comes from the 
pion self energy, thus we find that the value of $\alpha_{\rm S}$ is even 
smaller.

\subsubsection{Observation \,\char 35 1}
If $\alpha_{\rm S} \sim 0.3$, as this analysis of the N-$\Delta$
splitting suggested, the spin-orbit problem in the baryon spectrum 
is essentially resolved.

\subsubsection{Observation \,\char 35 2}
{\bf Q)} How can one ever seriously believe ``shell model'' type studies 
if effects of channel coupling are so large

\noindent{\bf A)} {\em You can't!}
Each resonance should be studied as a coupled channel scattering 
problem\ldots
Until recently this has been seen as involving too much work!  
So far it has only been taken seriously in a limited number of cases
\cite{Pearce:1986du,Schutz:1998jx} -- see also the discussion of  
Thomas and Miller\cite{Thomas:1991kz}.

\section{Conclusion}

In this very limited space we have tried to summarise a great deal of
physics that is not often taught. It is assumed that students of the
strong interaction have somehow ``picked up'' the information. In this
final, brief section we would like to make a few connections between
the simple models and ideas presented here and some modern research
topics.

In the past few years it has been realized that a hyperfine interaction
involving spin and isospin, rather than spin and colour (as for one
gluon exchange) offers some phenomenological improvements in fitting the
baryon spectrum \cite{Glozman:1996fu}. The motivation for such an interaction is
that there should be some residual short-distance remnant of the long
range pion cloud required by chiral symmetry that we presented here
using the CBM. Indeed, it has sometimes been argued that one should
include {\bf only} this interaction with {\bf no} one gluon exchange
term. 

There is currently great experimental activity and related theoretical
interest in exploring the chiral and deconfinement transitions expected
at high baryon density. One may ask whether these phase transitions
are coincident or even whether they exist at all. Recent work from
several different approaches to QCD\cite{Alford:1997zt,Bender:1997jf}, suggests that they may be
coincident, occurring at 3-4 times normal nuclear matter density. Moreover,
these calculations provide a rather unexpected motivation for a model of
hadron structure like the bag. Below the critical density for the
chiral/deconfining phase transition a region of space containing current
quarks is unstable and thus at finite, but low baryon density one
expects to form colourless bubbles of chirally restored phase -- just
like the MIT bag. There is still an enormous amount of work to be done
to move from this observation to a real understanding of the best way to
model the nucleon, but this work has provided important insight.

Chiral perturbation theory is currently enjoying tremendous popularity,
with many fascinating examples of experimental interest being explored
\cite{Bernard:1995dp,Donoghue:1998rp}. However, it has also been realized that the usual
formulation using, for example, dimensional regularization presents some
problems. In particular, the chiral corrections grow rapidly with the
mass of the Goldstone boson, whereas one would physically expect it to
be less important. The CBM has exactly the behaviour as chiral
perturbation theory as $m_\pi \rightarrow 0$, but the corrections to
hadron properties actually get smaller as the Goldstone boson mass
increases because of the physically motivated momentum cut-off in the
model -- recall this cut-off is related to the internal quark 
structure of the hadron. There is a great deal of important work to be
done to combine these two approaches to the same underlying physics --
e.g. see Ref. \cite{Donoghue:1998rp}.

With regard to the chiral structure of the nucleon we must mention the
very exciting experimental results from Fermilab in which, for the first
time, we have been given a direct view of the asymmetry between
$\bar{d}$ and $\bar{u}$ quarks in the nucleon sea \cite{Hawker:1998ty}. It seems
most likely that this asymmetry, which was anticipated on the basis of
the CBM \cite{Thomas:1983fh}, is at least in part due to the pion cloud of the nucleon
required by chiral symmetry \cite{Melnitchouk:1998rv}.

We cannot finish a review such as this without mentioning the very
fundamental questions at the heart of nuclear theory, for which models
such as those described here have an important role to play. In
particular, we would very much like to understand the role that nucleon
substructure plays in understanding the properties of finite nuclei as
well as matter at higher density. As just one example, the quark meson
coupling model \cite{Guichon:1988jp,Guichon:1996ue}, which is based on the MIT bag,
offers a completely new saturation mechanism for nuclear matter that
seems very natural. It also offers a consistent 
theoretical framework for calculating the
changes in hadron properties (masses, from factors and so on) in a
nuclear medium. There is a great deal of experimental interest in these
questions and we hope that these notes may be of some assistance in
understanding what is being done and perhaps in contributing to it.

\section*{Acknowledgments}
This work was supported in part by the Australian Research Council.

\section*{Appendix}

\subsection*{Dirac Matrices and Spinors}
\label{app:Math}
Throughout this paper we have followed the conventions of Bjorken and Drell \cite{Bjorken:1965dk}.
The gamma matrices satisfy the following condition :
\be
\{\gamma^{\mu},\gamma^{\nu}\} = \gamma^{\mu}\gamma^{\nu} + \gamma^{\nu}\gamma^{\mu} = 2g^{\mu\nu}.
\ee
Where the $\gamma$ matrices are defined by:
\be
\gamma^{0} = \beta,
\ee
\be
\vec{\gamma} = \beta\vec{\alpha},
\ee
\bea
\gamma_{5} = \gamma^{5} & = & i\gamma^{0}\gamma^{1}\gamma^{1}\gamma^{3} \non \\
& = & -i\gamma_{0}\gamma_{1}\gamma_{2}\gamma_{3},
\eea
and their values in the Dirac representation are:
\be
\gamma^{0} = \left( \begin{array}{cc}
I & 0 \\ 
0 & -I
\end{array} \right),
\ee
\be
\vec{\gamma} = \left( \begin{array}{cc}
0 & \vec{\sigma} \\
-\vec{\sigma} & 0
\end{array} \right),
\ee
\be
\gamma_{5} = \left( \begin{array}{cc}
0 & I \\
I & 0
\end{array} \right).
\ee
Obviously $\gamma^{0}$ is hermitian, $\gamma^{i}$ is antihermitian, $\gamma^{2}_{5} = I$, and finally
\be
\{\gamma_{5},\gamma^{\mu}\} = 0, \hspace*{15mm} \gamma^{\mu} = \gamma^{0}\gamma^{\mu\dagger}\gamma^{0}.
\label{eqn:gamma5_anticom}
\ee

\subsection*{Selected Solutions for Exercises}

\subsubsection*{Exercise 1}
We find the spin-flavour wave function of the $J_{z}=\half, \Delta^{+}$ in the following way.  We act with the spin lowering operator on each quark as shown below.
\beas
S_{-}\left|\Delta^{+},J_{z}=\frac{3}{2}\right> &  \propto 
& \left(S_{q}\right)_{-} \ket{\widetilde{u\up u\up d\up}} \\
& = \frac{1}{\sqrt{9}} & \left\{\ket{u\dn u\up d\up} + \ket{u\up u\dn d\up} + \ket{u\up u\up d\dn} + \right.\\
& & \left.\ket{u\dn d\up d\up} + \ket{u\up d\dn d\up} + \ket{u\up d\up d\dn} +\right. \\
& & \left.\ket{d\dn u\up d\up} + \ket{d\up u\dn d\up} + \ket{d\up u\up d\dn}\right\} \\
& = \frac{1}{3} & \left( \ket{\widetilde{u\up u\up d\dn}} + \ket{\widetilde{u\up u\dn d\up}}\right).
\eeas

\subsubsection*{ Exercise 2}
We want to find the wave function of
\bd
\ket{p\up} = \ket{uud,J_{z}=\half},
\ed
which is totally symmetric in Spin-Flavour.  We now find all totally symmetric combinations of these spin and flavour combinations.
\beas
\left(u\up u\up d\dn\right)_{\rm Symm(Spin-Flav)} = (u\up u\up d\dn) + (u\up d\dn u\up) \\
+ (d\dn u\up u\up)
\eeas
and
\beas
\left(u\up u\dn d\up\right)_{\rm Symm(Spin-Flav)} & = & (u\up u\dn d\up) + (u\dn u\up d\up) \\
& & (u\up d\up u\dn) + (u\dn d\up u\up) \\
& & (d\up u\up u\dn) + (d\up u\dn u\up)
\eeas
We now antisymmeterise with respect to, say, spin ($\up\dn$), and we get
\bd
\ket{p\up} = \frac{1}{\sqrt{18}}\left(2\ket{\widetilde{u\up u\up d\dn}} - \ket{\widetilde{u\up u\dn d\up}}\right)
\ed
where $18 = 2^{2}\times3+(-1)^{2}\times6$ (the factors 3 and 6 indicate the 
number of terms in the kets respectively). Alternatively, we can simply
write down the normalized combination of $\ket{\widetilde{u\up u\up
d\dn}}$ and $\ket{\widetilde{u\up u\dn d\up}}$ orthogonal to the 
$\Delta^{+},J_{z}=\frac{1}{2}$ state.

\subsubsection*{ Exercise 3}

We shall just present the solutions here, as the reader should be able to 
obtain these results alone by now.
\beas
\ket{\Sigma^{0}\up} & = & \frac{1}{\sqrt{18}}\ket{2(u\up d\up s\dn)_{\rm S} - (u\up d\dn s\up)_{\rm S} - (u\dn d\up s\up)_{\rm S}} \\
& = & \frac{1}{\sqrt{18}}\left(2\ket{\widetilde{u\up d\up s\dn}} - \ket{\widetilde{u\up d\dn s\up}} - \ket{\widetilde{u\dn d\up s\up}}\right) \\
\ket{\Lambda\up} & = & \frac{1}{\sqrt{6}}\left(\ket{\widetilde{u\up u\dn s\up}} - \ket{\widetilde{u\dn d\up s\up}}\right)
\eeas

\subsubsection*{ Exercise 4}

We know that 
\bd
\sum_{i=1}^{3}\frac{\mu_{iz}}{2} = \mu_{0}\left\{\frac{2}{3}(\vec{S}_{u})_{z} - \frac{1}{3}(\vec{S}_{d})_{z} - \frac{1}{3}(\vec{S}_{s})_{z}\right\}.
\ed
So we find
\beas
\bra{p\up}\sum_{i}\frac{\mu_{iz}}{2}\ket{p\up} & = &\frac{1}{18}\mu_{0} \bra{2(u\up u\up d\dn)_{\rm S} - (u\up u\dn d\up)_{\rm S}}\frac{2}{3}(\vec{S}_{u})_{z} \\
& &- \frac{1}{3}(\vec{S}_{d})_{z} - \frac{1}{3}(\vec{S}_{s})_{z}\ket{2(u\up u\up d\dn)_{\rm S} - (u\up u\dn d\up)_{\rm S}}\\
& = &\frac{\mu_{0}}{18} \left\{4\bra{(u\up u\up d\dn)_{\rm S}}\sum_{i}\mu_{iz}\ket{(u\up u\up d\dn)_{\rm S}}\right.\\
& &\left.- \bra{(u\up u\dn d\up)_{\rm S}}\sum_{i}\mu_{iz}\ket{(u\up u\dn d\up)_{\rm S}}\right\}.
\eeas
Therefore we find that
\beas
\bra{p\up}\sum_{i}\frac{\mu_{iz}}{2}\ket{p\up} = \frac{\mu_{0}}{18}\left\{4\cdot(\frac{2}{3})\cdot1\cdot3\right. & + & 4\cdot(-\frac{1}{3})\cdot(-\frac{1}{2})\cdot 3 \\
1\cdot(\frac{2}{3})\cdot0\cdot6 & + & \left.1\cdot(-\frac{1}{3})\cdot(\half)\cdot 6 \right\},
\eeas
where the factors in each of the terms of the sum are the factor multiplying the ket, the factor multiplying the spin operator, the sum of the spins of the relevant particle, and the number of terms in the ket, respectively.  This simplifies to
\bd
\mu_{p} \equiv \bra{p\up}\sum_{i}\mu_{iz}\ket{p\up} = \mu_{0}
\ed
Similarly we find that the wavefunction of the spin-up neutron is given by
\bd
\ket{n\up} \propto I_{-}\ket{p\up}
\ed
Therefore we find that $\mu_{n} = \frac{- 2 \mu_{0}}{3}$, 
and hence $\frac{\mu_{p}}{\mu_{n}} = -\frac{3}{2}$.

\subsubsection*{ Exercise 5}

Beginning with the fact that
\bd
\left(\vec{\lambda}_{1} + \vec{\lambda}_{2} + \vec{\lambda}_{3} \right)^{2}\left|N\right> = 0.
\ed
We fully expand the term that is squared.  We then use the fact that the particles are indistinguishable.  Thus we note that, for instance $\vec{\lambda}_{1}\cdot\vec{\lambda}_{2} \equiv \vec{\lambda}_{1}\cdot\vec{\lambda}_{3}$, so we find
\bd
3\vec{\lambda}^{2} + 3\vec{\lambda}_{1}\cdot\vec{\lambda}_{2} = 0,
\ed
or rearranging we get that
\bd
\vec{\lambda}_{1}\cdot\vec{\lambda}_{2} = -\vec{\lambda}^{2}.
\ed
Now, we use the hint that is given
\bd
\sum_{a}\left(\lambda_{a}\right)^{2} = \frac{16}{3} \equiv \vec{\lambda}^{2},
\ed
(the sum of the index $a$ is a sum from 1 to 8 -- due to the fact that we are working in SU(3)).  Substituting back be get
\beas
\left<\vec{\lambda}_{1}\cdot\vec{\lambda}_{2}\right>_{\rm Baryons} & = &\half\left<\vec{\lambda}_{1}\cdot\vec{\lambda}_{2}\right>_{\rm Mesons} \\
& = & \half\left<\vec{\lambda}^{2}\right> \\
& = & \half \left(-\frac{16}{3}\right) = -\frac{8}{3},
\eeas
as required.

\subsubsection*{ Exercise 9}

The requirement is to show that $[H,\vec{j}] = 0 = [H,K] = [\vec{j},K]$.  Since the working is similar in each case we shall only present a result for the last commutator.  We have that
\bd
[\vec{j},K] \propto \left[\vec{l}+\frac{\vec{\sigma}}{2},\vec{\sigma}\cdot\vec{l}\right],
\ed
where the proportionality comes from the fact that K is actually multiplied by $\gamma_{0}$.  Thus we find
\beas
[\vec{j},K] & \propto & \left[\vec{l},\vec{\sigma}\cdot\vec{l}\right] + \half\left[\vec{\sigma},\vec{\sigma}\cdot\vec{l}\right] \\
& = & \sigma_{j}[l_{i},l_{j}] + \half[\sigma_{i},\sigma_{j}]l_{j} \\
& = & \sigma_{j}i\varepsilon_{ijk}l_{k} + \half2i\varepsilon_{ijk}\sigma_{k}l_{j} \\
& = & i\varepsilon_{ijk}(\sigma_{j}l_{k} - \sigma_{j}l_{k}) \\
& = & 0,
\eeas
as required.

\subsubsection*{Exercise 11}

To show that $u$ and $l$ are continuous at $r=R$ we need to find solutions for them at $r\leq R$ and $r>R$ and then match at the boundary.  For the case of $r\leq R$, Eq. (\ref{eqn:cont_eqn}) simplifies to 
\bd
u'' + E^{2}u = 0
\ed
Solving this for $u$ we find that $u = A\sin(Er)$.  For $r>R$ we have
\bd
u'' + (E^{2}-m^{2})u = 0
\ed
with the general solution $u=\alpha\exp^{\sqrt{E^{2}-m^{2}}r}+\beta\exp^{-\sqrt{E^{2}-m^{2}}r}$.  Requiring that that $u$ is finite as the value of r increases (i.e. $u<\infty$ as $r\rightarrow\infty$) we find this general answer simplifies to $u=\beta\exp^{-\sqrt{E^{2}-m^{2}}r}$.

We now require that these solutions match at the boundary ($r=R$).  Thus we have
\beas
u(r=R) & = & A\sin(ER) \\
& = & \beta\exp^{-\sqrt{E^{2}-m^{2}}R},
\eeas
so we see the solution has the form
\bd
u = A\sin(ER)\exp^{-\sqrt{E^{2}-m^{2}}(r-R)}
\ed
Now we require that $f(r)$ is continuous at $r=R$.  In the simple case we are examining ($\kappa = -1$) we  see $f$ is given by
\bd
f=(E+(m+V_{\rm S}))^{-1}\frac{dg}{dr}
\ed
Again, using the substitution $g=\frac{u}{r}$ we have that
\bd
\frac{dg}{dr} = \frac{1}{r}\left(\frac{du}{dr}-\frac{u}{r}\right).
\ed
Substituting our two solutions for $u$ at the point $r=R$ we get
\bd
\frac{1}{E}\left(E\cos(ER)-\frac{\sin(ER)}{R}\right) = \frac{1}{E+m}\left(-\sqrt{m^{2}-E^{2}}\sin(ER) - \frac{\sin(ER)}{R}\right)
\ed
Rearranging, we find what we sought to find
\bd
\cos(ER) + \frac{\sqrt{1-(E/m)^{2}}}{1+(E/m)}\sin(ER) = \frac{\sin(ER)}{ER}\left(1-\frac{E}{E+m}\right).
\ed

\section*{References}
\nocite{*}                   
\bibliography{anu}        
\bibliographystyle{utphys}   

\end{document}